\documentclass[conference]{IEEEtran}
\IEEEoverridecommandlockouts
\usepackage{svg}
\usepackage{soul}
\usepackage{cite}
\usepackage{hyperref}
\usepackage{amsmath,amssymb,amsfonts}
\usepackage{graphicx}
\usepackage{textcomp}
\usepackage{xcolor}
\usepackage{caption}
\usepackage{adjustbox}
\usepackage{subcaption}
\usepackage{color}
\usepackage[left=0.680in, right=0.620in, bottom = 1in, top=0.7in]{geometry}
\usepackage{url,amssymb,threeparttable,multirow,booktabs, tabularx}
\def\BibTeX{{\rm B\kern-.05em{\sc i\kern-.025em b}\kern -.08em
    T\kern-.1667em\lower.7ex\hbox{E}\kern-.125emX}}
\usepackage{algorithm}
\usepackage{algpseudocode}
\usepackage{pifont}

\begin{document}

\title{Retrieval Augmented Generation with Multi-Modal LLM Framework for Wireless Environments}
\author{
    \IEEEauthorblockN{
         Muhammad Ahmed Mohsin\IEEEauthorrefmark{1}, Ahsan Bilal\IEEEauthorrefmark{2}, Sagnik Bhattacharya\IEEEauthorrefmark{1}, John M. Cioffi\IEEEauthorrefmark{1}
        }
       
   \IEEEauthorblockA{\IEEEauthorrefmark{1}Dept. of Electrical Engineering, Stanford University, Stanford, CA, USA}

   \IEEEauthorblockA{\IEEEauthorrefmark{2}Dept. of Computer Science, University of Oklahoma, OK, USA}

    \IEEEauthorblockA{Email: \{muahmed, sagnikb, cioffi \}@stanford.edu, ahsan.bilal-1@ou.edu
    }
}

\maketitle
\vspace{-40pt}

\begin{abstract}
Future wireless networks aims to deliver high data rates and lower power consumption while ensuring seamless connectivity, necessitating robust wireless network optimization. To achieve this, wireless network optimization is necessary. Large language models (LLMs) have been deployed for generalized optimization scenarios. To take advantage of generative AI (GAI) models, retrieval augmented generation (RAG) is proposed for multi-sensor wireless environment perception. Utilizing domain-specific prompt engineering, we apply retrieval-augmented generation (RAG) to efficiently harness multimodal data inputs from sensors in a wireless environment. Wireless environment perception is necessary for global LLM optimization tasks. Key pre-processing pipelines including image-to-text conversion, object detection, and distance calculations for multimodal RAG input from multi-sensor data from different devices are proposed in this paper to obtain a unified vector database crucial for optimizing large language models (LLMs) in global wireless tasks. Our evaluation, conducted with OpenAI’s GPT and Google’s Gemini models, demonstrates an 8\%, 8\%, 10\%, 7\%, and 12\% improvement in relevancy, faithfulness, completeness, similarity, and accuracy, respectively, compared to conventional LLM-based designs. Furthermore, our RAG-based LLM framework with vectorized databases are computationally efficient providing real time convergence under latency constraints.\footnote{This research is conducted in collaboration with Intel Corporation, Samsung Semiconductors, and Ericsson. The first two authors contributed equally to this work.}

\end{abstract}

\begin{IEEEkeywords}
Retrieval Augmented Generation, LLMs, Wireless Communication, Multi modal LLM, 6G.
\end{IEEEkeywords}

\section{Introduction}






Foundation models, developed by training extensively on large-scale unlabeled datasets to handle a variety of tasks, have become a fundamental pillar in contemporary AI research and its applications. These generative AI (GAI) models have led to cutting-edge applications such as ChatGPT and Gemini. They demonstrate remarkable generalization abilities, capturing complex relationships within data and large-scale neural network (NN) architectures with millions of parameters, enabling their application in every field. Large language models (LLMs), when well suited for dynamic environments, have the potential to revolutionize decision-making, resource management, and intelligent real-time optimizations for wireless networks~\cite{10847914}. 

To enhance the accuracy of these GAI models, RAG has been proposed to integrate factual information from external sources, creating a knowledge base for GAI models~\cite{NEURIPS2020_6b493230}. LLMs largely operate only on text based data, which restricts their usage for real time multi-modal functionalities. RAG-based services are currently popular for cloud deployment due to resource constraints~\cite{salemi2024optimization}. RAG enables LLMs to interpret diverse data modalities, which facilitate wireless network tasks like resource allocation and optimization. In the context of wireless networks, RAG-based knowledge bases can assimilate information from various sources, including network standards, research publications, and real-time sensor data providing multi-modal inference capabilities. Currently, RAG-based services in the radio access network (RAN) for internal RAN management and external B2B services in 6G are gaining popularity~\cite{huang2024toward}.

Next generation wireless networks are expected to offer high data rates, lower latency, and ubiquitous connectivity to support a diverse array of applications such as augmented reality, autonomous vehicles and telemedicine~\cite{celik2024dawn}. These networks will incorporate reconfigurable intelligent surfaces (RIS), Terahertz Communication (Thz), high altitude platforms (HAPs), terrestrial non-terrestrial networks (TN-NTNs). These complex technologies require complex systems for further resource optimization for their realization. With the use of large antenna arrays, channels have transitioned from being predominantly probabilistic to more deterministic~\cite{nazarenwar}, making sensing technologies crucial for achieving improved data rates and overall network performance in future wireless systems. Current techniques like reinforcement learning (RL), convex optimization face challenges regarding scalability, interpretability, and robustness. LLMs as general purpose optimizers are a potential solution to this problem~\cite{nie2024importance}, but for optimization in wireless networks, the LLMs need to be trained on multimodal sensory data for visual perception of the complete environment at each time step. RAG-based solution enable LLms for complete wireless environment perception through multimodal data input \cite{zou2024genainet}. Furthermore, fine-tuning LLMs for specific tasks can adapt LLMs for wireless network optimization, the computational overhead for fine-tuning is simply very large to adapt to dynamic environments~\cite{nazarenwar}.
Integrated Sensing and Communications (ISAC) has been proposed to incorporate Generative AI (GAI) into wireless communications, combining sensing and communication modules for efficient optimization via multimodal RAG perception. While some studies explore GAI in wireless applications, such as use case extraction \cite{mohsin2025deepreinforcementlearningoptimized}, \cite{erak2024leveraging} for 6G networks, substantial work on RAG-based multimodal input for wireless environment perception remains limited. This paper addresses this gap, expanding RAG applications for enhanced wireless optimization.


 \begin{figure*}[t!]
    \centering
    \includegraphics[width=1\linewidth]{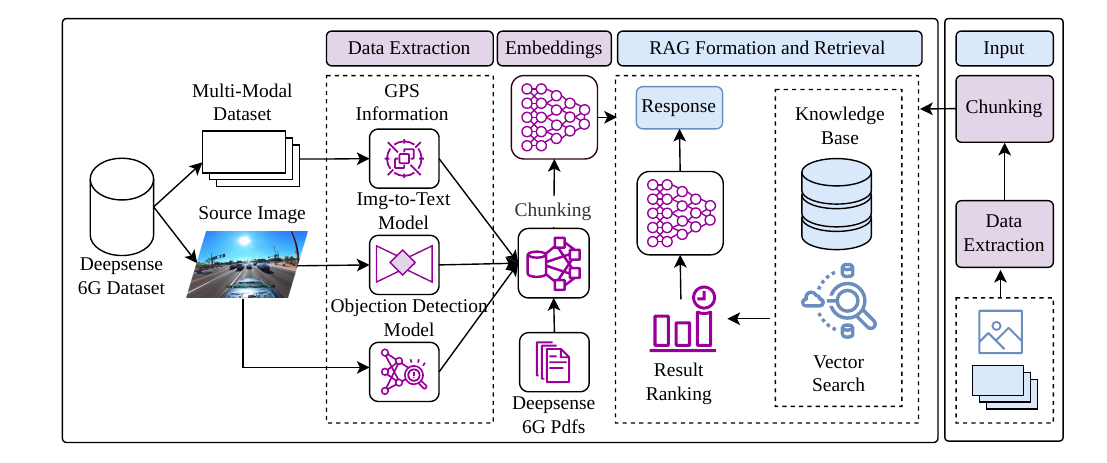}
    \caption{Work flow for vector database creation for RAG-based LLM wireless environment perception}
    \label{fig:workflow}
\end{figure*}

The key contributions of this paper are as follows: It leverages the DeepSense 6G dataset~\cite{alkhateeb2023deepsense6glargescalerealworld} for wireless environment perception through a multimodal RAG-based implementation. The DeepSense 6G dataset captures the complexities of real-world wireless environments, incorporating a wide range of real-world effects. Camera images are converted to textual data using an image to text model, and GPS detects the location of the transceiver, and Lidar is used to perceive the vehicles and objects in the image using YOLO~\cite{jiang2022review}. This is all converted to textual information for the LLM model and a knowledge base is created using ChromaDB~\cite{xie2023brief}. RAG-based, contextually aware, wireless environment perception is then tested on several open-source language models to observe the performance in terms of relevance, contextual recall, correctness, and faithfulness. Results demonstrate that RAG-based environment perception is essential for resource optimization tasks, and RAG-based models perform better than baseline methods by 8\%, 8\%, 10\%, 7\%, and 12\% in terms of relevancy, faithfulness, completeness, similarity, and accuracy.
 
\section{Deepsense 6G Dataset}
\subsection{Environment Description} This work use the DeepSense 6G dataset with scenario 36. The system model consists of an outdoor wireless environment with high frequency V2V communication. The total number of samples is 112,200 with scenario 36 having 24,800 samples. Scenario 36 is modeled as inter-city scenario with long drives and the data modalities include 360° RGB images, 4 x 64-dimensional received power vector, GPS locations and 3D LiDAR point-cloud.  For the data collection, the units used are mobile transmitter, mobile receiver, wireless sensor, visual sensor, position sensor and 3D Lidar sensor. The mobile transceiver contains four 60-GHz mmWave Phased arrays, 360-degree RGB camera, 3D LiDAR (32 vertical, 1024 azimuth channels) and a GPS kit. The visual camera for RGB images is 360-degree camera with 5.7k resolution. Furthermore, GPS tracker has horizontal accuracy within one meter. The 3D Lidar sensor is equipped with 32 thousand-point 3D point cloud and 20Hz maximum motor spin frequency.
\subsection{Data Acquisition} The data preprocessing involves two steps. The first step is converting sensor data to timestamped samples and filtering, organizing, and synchronizing data. The second step is data visualization, which uses DeepSense library~\cite{alkhateeb2023deepsense6glargescalerealworld} to render scenario videos with synchronized data.

\section{ Proposed Framework: Multi-Modal RAG for Wireless Environment Perception}
\subsection{Multi Modal Data Preprocessing}
\subsubsection{Text to image description} LLM models are restricted to textual descriptions for the input dataset, as depicted in Fig.~\ref{fig:workflow}. To incorporate the diversity of sensor data, we first uses  GPT-4o \cite{openai2023gpt4} for image-to-text description of the dataset's RGB images. The prompt for text-to-image description describes all the image's elements and the scenery present in detail. GPT-4o is trained on a single new model end-to-end across text, vision, and audio, meaning that all inputs and outputs are processed by the same neural network. This is the first step for the knowledge base creation.
\subsubsection{GPS calculations} Within data extraction, a GPS sensor provides longitude, latitude, and altitude for transceiver locations. For directional bearing angles, data extraction provides longitude and latitude values only. 
The distance between two points on the Earth's surface, given by their latitudes and longitudes, can be calculated using the Haversine formula:
\begin{equation}
    d = 2R \cdot \arctan2\left( \sqrt{a}, \sqrt{1 - a} \right)
\end{equation}
\begin{equation}
    a = \sin^2\left(\frac{\Delta \phi}{2}\right) + \cos(\phi_1) \cdot \cos(\phi_2) \cdot \sin^2\left(\frac{\Delta \lambda}{2}\right)
\end{equation}
\begin{equation}
    \Delta \phi = \phi_2 - \phi_1 \quad \text{and} \quad \Delta \lambda = \lambda_2 - \lambda_1
\end{equation}
where \( \phi_1 \) and \( \phi_2 \) are the latitudes of points 1 and 2 in radians, and \( \lambda_1 \) and \( \lambda_2 \) are the longitudes in radians. \( R \) is the Earth's radius (in kilometers). The initial bearing angle between two GPS coordinates ($ \theta_1, \theta_2$) is given by:
\begin{equation}
    \theta_1 = \sin(\Delta \lambda) \cdot \cos(\phi_2)
\end{equation}
\begin{equation}
    \theta_2 = \cos(\phi_1) \cdot \sin(\phi_2) - \sin(\phi_1) \cdot \cos(\phi_2) \cdot \cos(\Delta \lambda)
\end{equation}
where \( \Delta \lambda = \lambda_2 - \lambda_1 \), and \( \phi_1 \), \( \phi_2 \), \( \lambda_1 \), and \( \lambda_2 \) are the latitudes and longitudes (in radians) of points 1 and 2. The compass bearing (normalized to the range 0–360°) is given by:
\begin{equation}
    \text{Bearing} = (\theta \cdot \frac{180}{\pi} + 360) \mod 360,
\end{equation}
where \( \theta \) is the initial bearing in radians. The text-to-image function incorporates these modalities into the modal.


\subsubsection{Object detection using YOLO} Data extraction uses the YOLO (You Only Look Once) object detection model for efficient localization and classification in RGB images. YOLO’s use of a “focal loss” function enhances its ability to detect small objects, outperforming standard cross-entropy loss, and processes images at up to 155 frames per second, making it faster than other state-of-the-art models. For improved perception of the wireless environment, the detected vehicle count in each image is also added to the image-to-text description.
\subsubsection{Lidar Images} Lidar (not shown but part of data extraction) provides 3D images for further information regarding the environment. These images are again pre-processed through image-to-text description models and help in the recognition of pedestrians and vehicles to enhance the perception.

\subsection{DeepSense 6G data retrieval}
To further enhance the knowledge base creation as in Fig.~\ref{fig:workflow}, the embedding's function web scrapes to gather all DeepSense 6G related documents and published research articles. First, the text data is extracted from these Pdfs using pytesseract. Then text chunking is performed using LangChain to make it manageable for the LLM model. Then tokenization is performed, converting each word to a token. The next step involves Chroma DB using various embedding models to convert text into vector embeddings. The embedding function uses transformer model \texttt{all-MiniLM-L6-v2}. One of the simplest forms of text vectorization is one-hot encoding. For a vocabulary of size V, each word is represented as a V-dimensional vector where all elements are 0 except for a single 1 at the index corresponding to that word. Chroma DB stores these embeddings along with the original text and the metadata. It uses efficient data structures to enable fast similarity search (semantic search) over these vectors. 
\subsection{Knowledge Base Creation with ChromaDB} Knowledge base creation with ChromaDB involves several steps. The first step involves creating a collection, which is analogous to a table in relational databases. This collection serves as the container for documents and their associated embeddings through an embedding model. The raw text undergoes preprocessing, which includes tokenization, lowercasing, and removal of stop words. The selected embedding model converts each preprocessed document into a high-dimensional vector representation. These vectors capture the semantic meaning of the text in a numerical format. The system associates the stored vectors with their corresponding metadata and unique IDs, allowing for efficient retrieval and filtering later. To enable fast similarity searches, Chroma DB builds an index over the stored vectors. This index is based on Approximate Nearest Neighbors (ANN) to facilitate rapid retrieval of similar vectors. 
\subsection{Ranking the results and LLM integration}
\subsubsection{Wireless Communications Domain-Specific Prompt Engineering for Multi-Modal Input} 
{Wireless domain knowledge is vital in designing prompts that precisely capture essential data, enhancing retrieval-augmented generation (RAG) for wireless optimization. Input data, including images and GPS coordinates, are processed to extract bearing angles, distances, and using image-to-text \cite{openai2023gpt4} and object count using object detection \cite{redmon2016lookonceunifiedrealtime}, making these prompts valuable assets for future wireless applications. This combined information is then structured to generate a prompt that aligns seamlessly with the RAG architecture. The prompt is structured by transforming its components and any real-time multi-modal sensory data into a unified and standardized format suitable for large language models (LLMs). This ensures the prompt is aligned with the knowledge base, enabling a higher cosine match during the RAG retrieval process, and the raw textual data of the matched use cases are then extracted from the regular database to compose the final prompt context.

 \begin{figure*}[t!]
    \centering
    \includegraphics[width=1\linewidth]{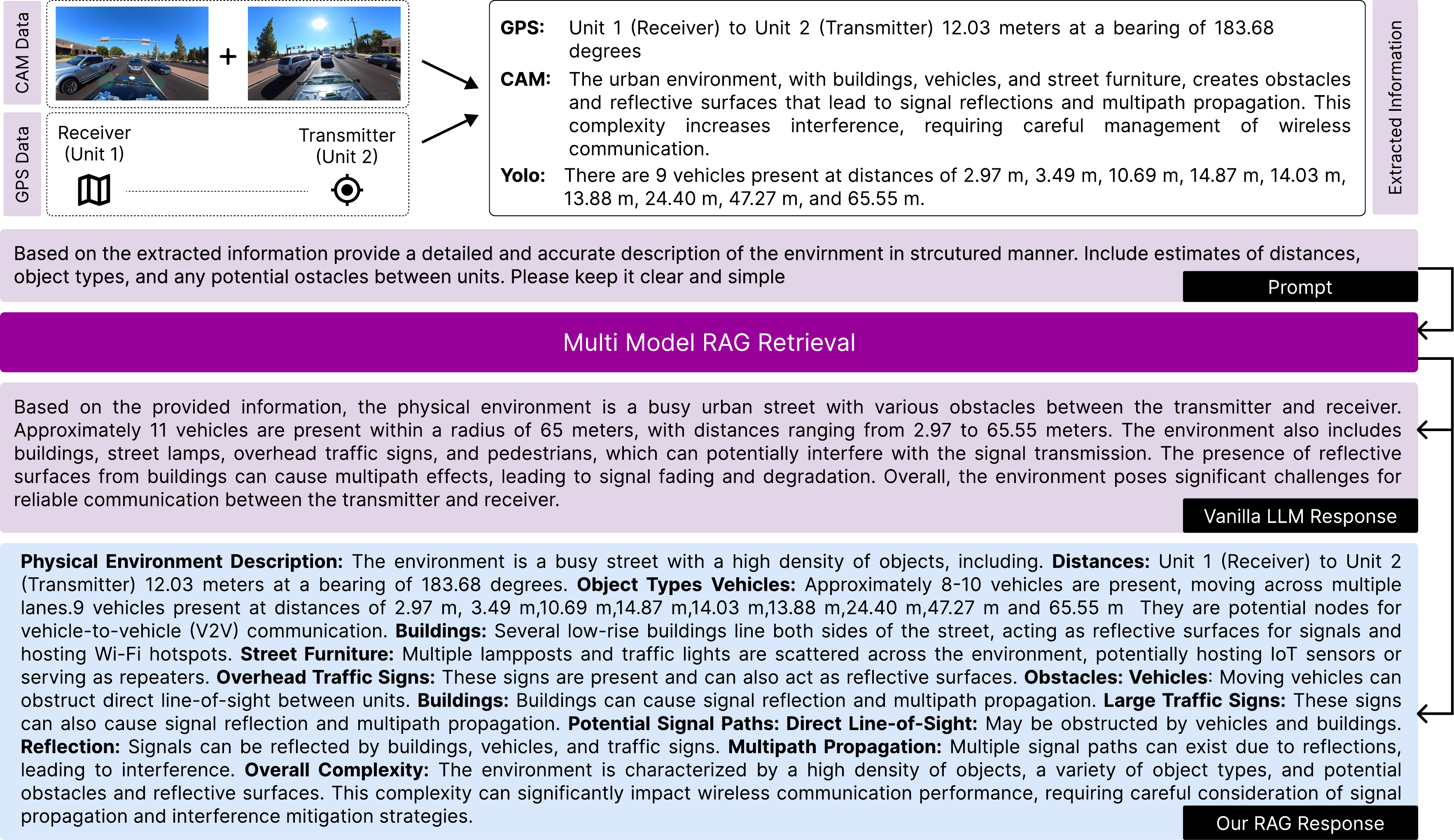}
    \caption{A detailed response comparison for RAG-based LLM vs. Vanilla LLM [baseline] }
    \label{fig:enter-label}
\end{figure*}

\subsubsection{Structured Information and result ranking} The structured prompt is further converted into numerical embeddings, allowing for efficient processing and comparison with the vectorized data in the knowledge base. 
The retrieval module uses this to gather relevant context from the knowledge base by doing the rank-based vector semantic search in the chromaDB \cite{xie2023brief} based knowledge base. The vector first used by the paper \cite{Sarmah2024HybridRAG} prioritizes the most contextually appropriate portions of the knowledge base. This refined search mechanism optimizes retrieval by focusing on the most relevant content. This context is then combined with a structured prompt to generate a control action. The framework incorporates vectorized databases which are computationally efficient and provides complete environment descriptions near optimal to coherent channel times for resource allocation problems.
\section{Quantitative and Qualitative Results}
Results are presented for two cases. \textbf{Case 1} evaluates RAG-based model narratives using accuracy and fluency metrics (Fig.~\ref{fig:performance_metric}) within the framework of \cite{zytek2024llms} (Fig.~\ref{fig:enter-label}). \textbf{Case 2} involves numerical analysis by refining ground truth text for all scenes in the knowledge base, addressing missing details and correcting extracted information. The initial knowledge base uses 360 scenes from the DeepSense6G dataset, as object types and blockages lack predefined labels. Testing was then conducted by randomly selecting scenarios and creating the corresponding ground truth. Results compare the performance of different vanilla LLM models with RAG in terms of faithfulness, correctness, and similarity score, using different LLMs as the retrieval component in RAG, as shown in Figure~\ref{fig:2}. This figure also assesses the contribution of different sections of the extracted information in the knowledge base and how these features influence the relevance of the RAG-generated output compared to the ground truth. Furthermore, results examine the impact of various large language models used in the RAG framework on relevance, as illustrated in Figure~\ref{fig:relevancy_comparison}.
The overall structure of the process is shown in Figure \ref{fig:enter-label}, where we compare the response of our RAG model with that of a vanilla LLM. Both models were tasked with generating a detailed description of a busy city street scene. To provide more descriptive insights, we selected a specific scenario featuring congested traffic, including cars, motorcycles, and other stationary objects along the road. Our RAG model delivers precise spatial analysis, identifying obstacles, assessing line-of-sight, and suggesting maneuvers.



\begin{figure*}[t!]
     \centering
    \begin{subfigure}[t]{0.66\columnwidth} 
         \centering
         \includegraphics[width=\textwidth]{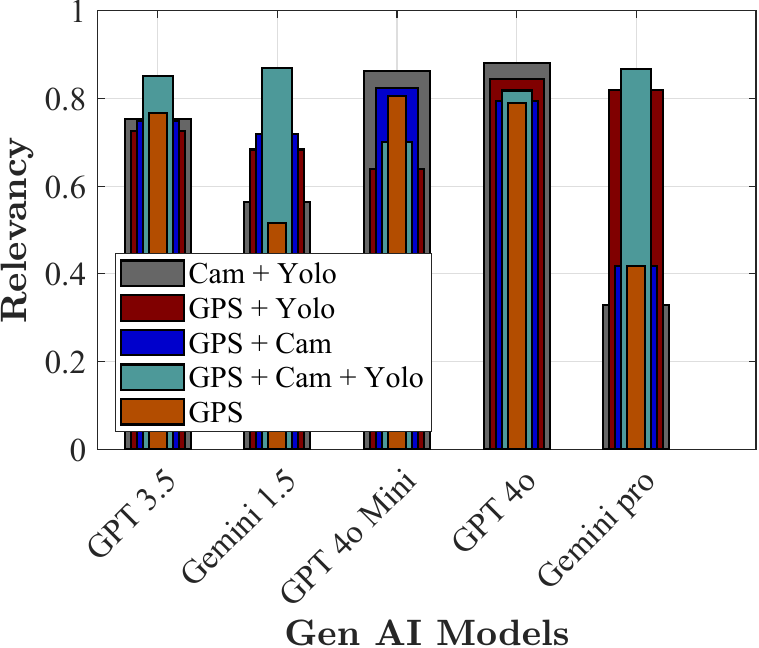}
         \caption{Gen AI models vs. relevancy}
         \label{fig:relevancy_comparison}
     \end{subfigure}
     \hspace{-0.01\columnwidth} 
     \begin{subfigure}[t]{0.66\columnwidth} 
         \centering
         \includegraphics[width=\textwidth]{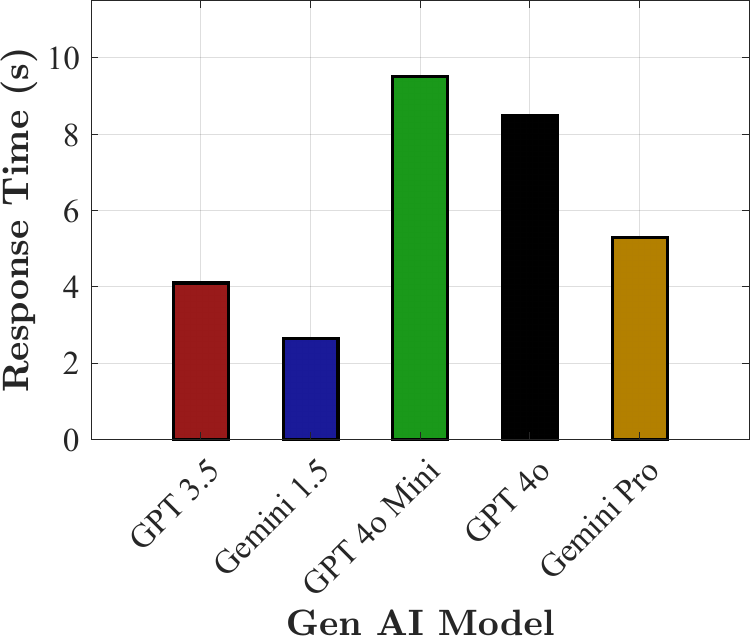}
         \caption{Gen AI models vs. response time [s]}
         \label{fig:reward}
     \end{subfigure}
     \hspace{-0.01\columnwidth} 
     \begin{subfigure}[t]{0.66\columnwidth} 
         \centering
         \includegraphics[width=\textwidth]{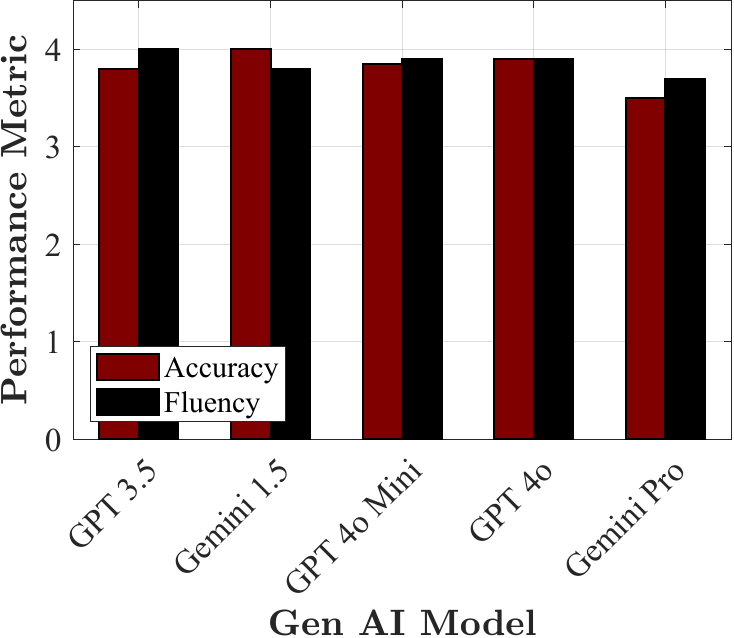} 
         \caption{Gen AI models vs. performance metric}
         \label{fig:performance_metric}
     \end{subfigure}
      \caption{
\textbf{(a)} Relevancy: RAG-based LLMs [Cam, YOLO, GPS]; \textbf{(b)} Response time: Gen AI models; \textbf{(c)} Accuracy \& fluency: RAG-based LLMs [0-4].}
     \label{fig:2}
\end{figure*}

\begin{figure*}[t!]
     \centering
    \begin{subfigure}[t]{0.66\columnwidth} 
         \centering
         \includegraphics[width=\textwidth]{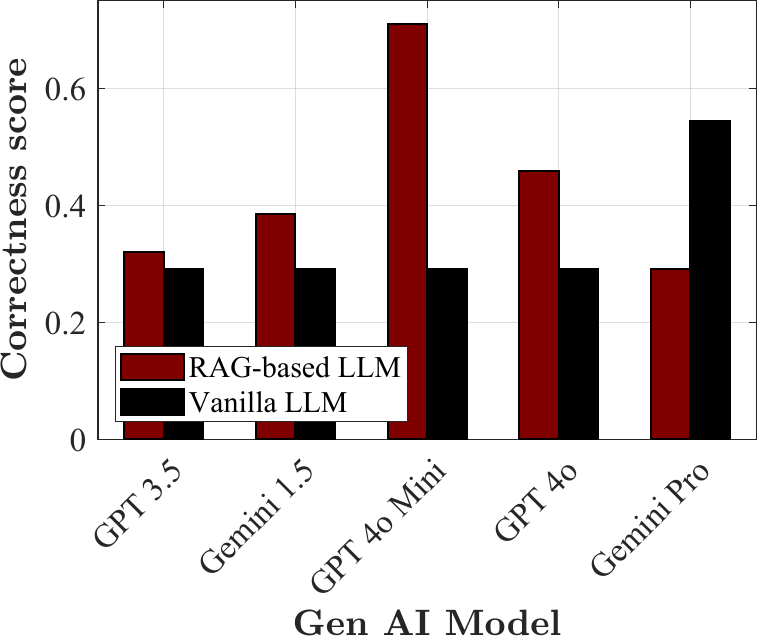}
         \caption{Gen AI models vs. correctness score.}
         \label{fig:Sumrate}
     \end{subfigure}
     \hspace{-0.01\columnwidth} 
     \begin{subfigure}[t]{0.66\columnwidth} 
         \centering
         \includegraphics[width=\textwidth]{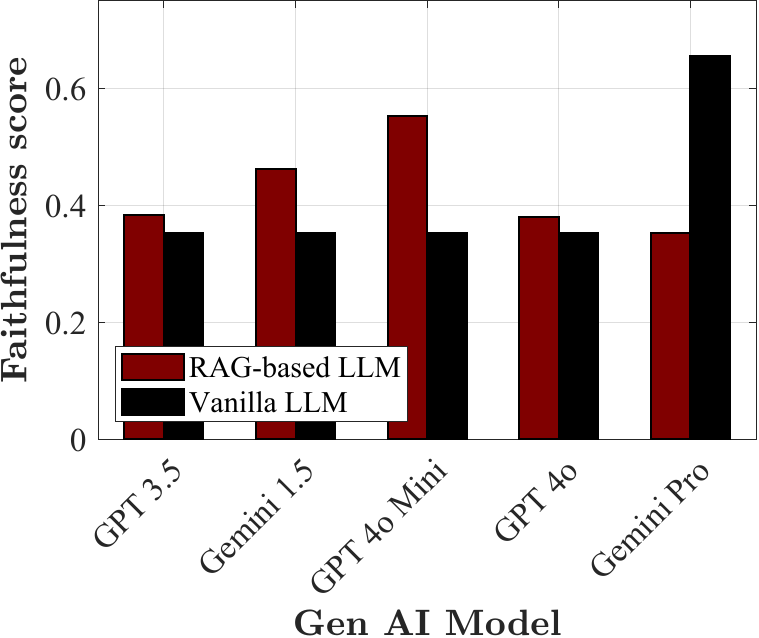}
         \caption{Gen AI models vs. faithfulness score}
         \label{fig:reward}
     \end{subfigure}
     \hspace{-0.01\columnwidth} 
     \begin{subfigure}[t]{0.66\columnwidth} 
         \centering
         \includegraphics[width=\textwidth]{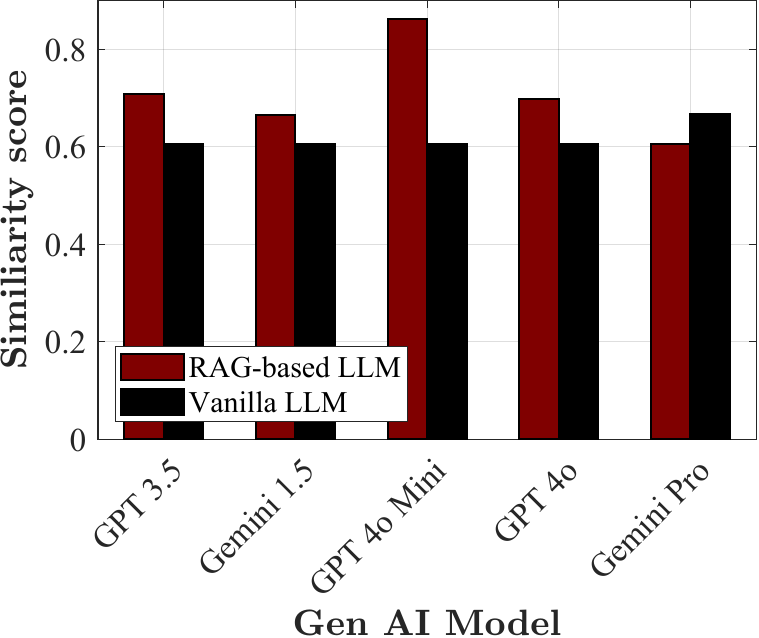} 
         \caption{Gen AI models vs. similarity score}
         \label{fig:outage}
     \end{subfigure}
      \caption{
\textbf{(a)} Correctness, \textbf{(b)} Faithfulness, \textbf{(c)} Similarity: RAG-based vs. baseline LLMs [0-1].}
     \label{fig:2}
\end{figure*}

\subsection{Evaluating Case 1:}
Simulations generated SHAP (shapley additive explanations) explanations on the output responses from our RAG model and selected the top three most important contributing features. The selected contributing features and the generated response from RAG were subsequently fed into the framework developed by \cite{zytek2024llms} to evaluate accuracy and fluency.
\subsubsection{Evaluation rubrics and numerical results}
In \cite{zytek2024llms}, model outputs were scored from 0 (lowest) to 4 (highest) based on fluency and accuracy. Fluency assessed naturalness, with 0 being "very unnatural" and 4 "very natural." Accuracy measured factual correctness, with 0 for outputs with errors and 4 for error-free outputs. This scoring system aimed to ensure that both the linguistic quality and the factual correctness of the model’s outputs were rigorously evaluated.
To assess RAG performance with different LLM models, figure \ref{fig:performance_metric} shows the results. These indicate that the \texttt{Gemini 1.5 flash} \cite{georgiev2024gemini} model slightly outperforms GPT-4.0 \cite{openai2023gpt4} in terms of accuracy, but GPT-4.0 exhibits superior fluency.

The results highlight that while the \texttt{Gemini} model has a marginal edge in accuracy, GPT-4.0 consistently generates more fluent narratives, demonstrating a trade-off between accuracy and fluency across different LLM models used in our RAG framework for wireless environments.
\subsection{Evaluating Case 2: }

\subsubsection{Correctness}
Results compared the responses generated by our RAG framework to those produced by a vanilla LLM, evaluating them in terms of correctness. The correctness was measured using a combination of cosine similarity and F1 score based on token overlap. Given two responses, the overall correctness \(\text{Correctness}\) is calculated using the following formula:
\begin{equation}
    \text{Correctness} = \omega \cdot \text{Cosine Similarity} + (1 - \omega) \cdot \text{F1 Score},
\end{equation}
where $\omega$ is a weighting factor, set to 0.25 in our experiments, allowing a balance between cosine similarity and the F1 score. \textbf{Cosine Similarity} measures the semantic similarity between the two texts by comparing their vector representations. \textbf{F1 Score} is based on token overlap between the generated response and the ground truth, balancing precision and recall.
\begin{equation}
    \text{cosine similarity}(r, g) = \frac{r \cdot g}{\|r\| \|g\|}
\end{equation}
\begin{equation}
    \text{precision} = \frac{|r \cap g|}{|r|}
\end{equation}
\begin{equation}
    \text{recall} = \frac{|r \cap g|}{|g|}
\end{equation}
\begin{equation}
    \text{F1 Score} = 2 \cdot \frac{\text{precision} \cdot \text{recall}}{\text{precision} + \text{recall}}
\end{equation}
Eq. (8) defines Cosine Similarity as the normalized dot product of response \( r \) and ground truth \( g \). Eqs. (9) and (10) define Precision and Recall, measuring token overlap in \( g \) and \( r \), respectively, while Eq. (11) gives the F1 Score, their harmonic mean. Overall correctness combines Cosine Similarity and F1 Score with weighted contributions, showing our RAG-based approach outperforms the vanilla LLM, as illustrated in Fig.~\ref{fig:Sumrate}.

\subsubsection{Faithfulness}
In addition to correctness, tests also evaluated the faithfulness of the responses generated by the RAG model compared to the vanilla LLM. Faithfulness measures how closely the generated output adheres to the ground truth by assessing the overlap of tokens between the result and the reference text. The faithfulness score is calculated as the ratio of common tokens between the result and the ground truth to the total number of tokens in the result. Mathematically:
\begin{equation}
 \text{faithfulness}(r, g) = \frac{|r \cap g|}{|r|},
\end{equation}
where $r$ represents the set of tokens in the generated response. $g$  represents the set of tokens in the ground truth. $|r \cap g|$ denotes the number of tokens common to both. This metric indicates how accurately the generated response reflects the reference text, with higher faithfulness suggesting more relevant and precise information from the ground truth. We used this faithfulness score, alongside the correctness metric discussed earlier, to further assess the quality of the RAG model's outputs. As shown in Figure~\ref{fig:reward}, the RAG framework generally produces more faithful responses than the vanilla LLM model.

\subsubsection{Semantic Similarity}
To assess the semantic similarity between the RAG-generated responses and the ground truth, results use sentence embeddings through the \texttt{SentenceTransformer} model, specifically the \texttt{all-MiniLM-L6-v2} variant from the \texttt{sentence-transformers} library. The \texttt{SentenceTransformer} efficiently converts sentences into dense vector embeddings, capturing semantic meaning for text comparison. The semantic similarity score, derived from the cosine similarity of the embeddings between the generated response and the ground truth, measures semantic closeness while accommodating minor wording differences. Ranging from -1 (dissimilar) to 1 (highly similar) and reflects sentence alignment that token-level comparisons might miss.

Mathematically, the cosine similarity between the two sentence embeddings \(e_r\) and \(e_g\) (for the result and ground truth, respectively) is given by:
\begin{equation}
    \text{cosine similarity}(e_r, e_g) = \frac{e_r \cdot e_g}{\|e_r\| \|e_g\|},
\end{equation}
where $e_r$  is the embedding vector for the generated response. $e_g$ is the embedding vector for the ground truth. $\|e_r\|$ and  $\|e_g\|$ are the magnitudes (norms) of the respective vectors. The results, in figure~\ref{fig:outage}, show that our RAG framework consistently produces responses with high semantic similarity to the ground truth, surpassing the vanilla LLM model in most cases.
\subsubsection{Real World Applicability}
The DeepSense 6G dataset incorporates real world sensor data captured in real time to train the RAG-based LLM for wireless environment perception. Due to vectorized databases, the prompt generation and perception is performed within seconds to comply with wireless network latency. RAG-based LLM models do not hallucinate much in comparison to fine-tuned models due to vectorized search spaces for answers. Wireless environment perception using RAG-based LLMs provides bases for real time resource allocation problem that current wireless standards face.

\section{Conclusion and Future Work}
In this paper, we present a comprehensive pre-processing pipeline for multimodal multisensory input data for wireless environment perception through RAG-based LLM models. For LLMs to optimize complex wireless environments, environment perception is necessary and due to multimodal output for each sensor in the environment a standard pre-processing pipeline is necessary. RAG-based LLM optimization for context-aware pipeline achieves 8\%, 8\%, 10\%, 7\%, and 12\% improvements in relevancy, faithfulness, completeness, similarity, and accuracy over baseline LLMs. For future work, this preprocessing pipeline can be integrated into ISAC systems to cater multisensory data for wireless environment optimization.

\footnotesize
\bibliographystyle{IEEEtran}
\bibliography{main.bib}

\end{document}